%% file: main.tex
\documentclass[10pt,a4paper]{article}

\input{preamble}

\usepackage{titling}
\title{Fast and accurate solvers for simulating Janus particle suspensions in Stokes flow}

\author{Ryan Kohl\thanks{Department of Mathematics, University of Michigan}, \, Eduardo Corona\thanks{Department of Mathematics, New York Institute of Technology} \ \thanks{Corresponding author. \emph{Email address:} ecorona@nyit.edu, ecorona.villanova@gmail.com} ,\ Vani Cheruvu\thanks{Department of Mathematics and Statistics, University of Toledo},\, and Shravan Veerapaneni\textsuperscript{*}}
\date{}

\begin{document}
\maketitle

\vspace{-4em}

\begin{abstract}
 We present a novel computational framework for simulating suspensions of rigid spherical Janus particles in Stokes flow. We show that long-range Janus particle interactions for a wide array of applications may be resolved using fast, spectrally accurate boundary integral methods tailored to polydisperse suspensions of spherical particles. These are incorporated into our rigid body Stokes platform. Our approach features the use of spherical harmonic expansions for spectrally accurate integral operator evaluation, complementarity-based collision resolution, and optimal $\mathcal{O}(n)$ scaling with the number of particles when accelerated via fast summation techniques. We demonstrate the flexibility of our platform through three key examples of Janus particle systems prominent in biomedical applications: \emph{amphiphilic, bipolar electric} and \emph{phoretic} particles. We formulate Janus particle interactions in boundary integral form and showcase characteristic self-assembly and complex collective behavior for each particle type.
\end{abstract}

\section{Introduction}
\label{sec:intro}
\input{introduction}

\section{Mathematical Preliminaries}
\label{sec:background}
\input{background}

\section{Discretization}
\label{sec:Discretization}
\input{Discretization}

\section{Amphiphilic Particles}
\label{sec:Amphiphilic}
\input{amphiphillic.tex}

\section{Bipolar Electric Particles}
\label{sec:Bipolar}
\input{Bipolar.tex}

\newpage
\section{Phoretic Particles}
\label{sec:Phoretic}
\input{phoretic.tex}

\section{Conclusions}
\label{sec:Conclusions}
\input{conclusions.tex}

\section{Acknowledgements} 
We acknowledge support from NSF under grants DMS-1719834, DMS-1454010 and DMS-2012424, and the Mcubed program at the University of Michigan (UM). Authors also acknowledge the computational resources and services provided by UM's Advanced Research Computing.
 
\appendix
\begin{appendices}
\input{Appendix.tex}
\end{appendices}

\printbibliography
\end{document}

%% file: preamble.tex
\usepackage[dvips]{color}
\usepackage [autostyle, english = american]{csquotes}
\usepackage[
pdftitle={BIE Janus particles},
pdfcreator={pdflatex},
pdfsubject={preprint},
hyperindex = {true},
colorlinks = {true},
linkcolor = {blue},
pagecolor = {blue},
citecolor = {blue}
]{hyperref}
 
\usepackage[backend=bibtex,
style=numeric,
sorting=none,maxbibnames=99]{biblatex} 
\DeclareUnicodeCharacter{0301}{KOKO}
\bibliography{bibliography}
\renewbibmacro{in:}{%
  \ifentrytype{article}
    {}
    {\bibstring{in}%
     \printunit{\intitlepunct}}}

\usepackage{tikz}
\usetikzlibrary{calc,trees,positioning,arrows,chains,shapes.geometric,  decorations.pathreplacing,decorations.pathmorphing,shapes,  matrix,shapes.symbols}
\usetikzlibrary{shapes,decorations}
\usepackage[english]{babel}
\usepackage{indentfirst}
\usepackage{amsmath}
\usepackage{algorithm,algpseudocode}
\usepackage{wrapfig}
\usepackage{bm}
\usepackage{amssymb}
\usepackage{caption}
\usepackage{booktabs}
\usepackage{authblk}
\usepackage{graphicx}
\usepackage{amsthm}
\usepackage{comment}
\newtheorem{theorem}{Theorem}
\newtheorem{lemma}{Lemma}
\usepackage[toc,page]{appendix}
\usepackage{float}
\usepackage[lofdepth,lotdepth]{subfig}

\DeclareMathAlphabet{\mathpzc}{OT1}{pzc}{m}{it} 
\usepackage{xargs}
\usepackage{matlab-prettifier}
\usepackage[nodisplayskipstretch]{setspace}
\numberwithin{equation}{section}
\newcommand{\nd}[1]{\frac{\partial #1}{\partial\nu}}

\newcommand{\norm}[1]{\|#1\|}
\newcommand{\SL}{\mathcal{S}}
\newcommand{\DL}{\mathcal{D}}
\newcommand{\8}{\infty}
\newcommand\pcal[1]{\mathpzc{#1}}
\newcommand{\SLStk}{\pcal{S}}
\newcommand{\DLStk}{\pcal{D}}
\newcommand{\TStk}{\pcal{K}}
\newcommand{\mbx}{\boldsymbol{x}}
\newcommand{\mbf}{\boldsymbol{f}}
\newcommand{\mbu}{\boldsymbol{{u}}}

\usepackage{wrapfig}

%% file: introduction.tex
\indent Although the term ``Janus particle" originated in the study of two-faced amphiphilic structure, it is now applied to a wide class of colloidal particles with more than a single type of surface chemistry or composition. The anisotropic structure in most Janus particles involves two hemispheres that differ in electric, magnetic or optical properties or in their physicochemical interaction with the surrounding fluid. Dense suspensions of Janus particles have been widely demonstrated to display complex aggregate behavior, clustering and self-assembly into larger-scale structures \cite{yang2012janus,walther2008janus,hong2008clusters}. Spurred by advances in design and manufacturing, the study of self-assembling materials based on Janus particle suspensions has garnered great interest, showing particular potential in biomedical applications such as drug delivery, medical imaging and manufacturing of biosensors and micromotors \cite{su2019janus,yang2012microfluidic,patra2013intelligent}.

In the study of dense particulate systems, direct numerical simulation (DNS) can play a crucial role to gain insight into their complex behavior and make accurate predictions. DNS of such systems, however, requires overcoming several numerical challenges: methods must accurately resolve short-ranged interactions (e.g., collisions) and computationally-intensive long-ranged, many-body interactions (e.g., hydrodynamic). Due to slow relaxation times and non-linearity typical of soft matter systems, robust and scalable solvers are required to tackle the large-scale, long-term simulations involved. 

Further, DNS of Janus particle suspensions epitomizes the demanding nature of multiphysics active matter systems. The rheology and motility of Janus particle suspensions, by their very essence, result from the coupling of one or several long-range physicochemical fields with the fluid flow. At particle surfaces, boundary conditions such as surface traction balance, induced fluid slip, and electromagnetic field jump conditions must be accurately satisfied. Additional coupling can occur in the bulk, for instance, due to advection of chemical solutes. Since self-assembly and clustering are the key phenomena of interest, resolving particle collisions and confinement effects are unavoidable. 

In part owing to these challenges, DNS studies of the hydrodynamics of Janus particle suspensions are rather scarce; fluid dynamics of a moderate number of particles have been studied in two \cite{fu2019amphiphilic,kanso2019phoretic,kim2017numerical,sobecki2021dynamics} and three spatial dimensions \cite{daghighi20113d}, while large-scale simulation studies have focused on statistical molecular simulation methods such as Molecular Dynamics, Lattice Boltzmann and Monte Carlo methods \cite{molotilin2016electrophoresis,baran2020self,banik2021substrate}. In this work, we seek to bridge this gap and propose an efficient, fast algorithmic framework for DNS of dense Janus suspensions in three dimensions, building on recent advances in the state-of-the-art rigid body Stokes solvers.  

Several numerical methods have been developed in the past few decades for simulating rigid particle suspensions. They may be classified according to how they resolve hydrodynamic particle interactions and near-field lubrication effects. These include approximation methods such as Rotne–Prager–Yamakawa \cite{wajnryb2013generalization}, Stokesian dynamics \cite{brady1988stokesian} and multipole methods \cite{cichocki1994friction}, as well as schemes that directly solve the underlying partial differential equation (PDE) such as fictitious domain methods, immersed boundary methods and boundary integral methods \cite{maxey2017simulation}. In the context of numerical PDE solution, boundary integral methods are particularly attractive due to dimensionality reduction and improved numerical conditioning. The key numerical hurdle for effective implementation of these methods is often the fast and accurate evaluation of boundary layer potentials. 
Overcoming this hurdle, recent contributions in this field have led to the implementation of large-scale simulation platforms for particulate Stokes flow simulation in three dimensions \cite{rahimian2010petascale, malhotra2015pvfmm, lu2018parallel, yan2019scalable}. 

In particular, a scalable computational framework adaptable to a large class of the Stokes mobility solvers was proposed in \cite{yan2019scalable} by incorporating a parallel complementarity formulation based collision resolution algorithm. In addition, a fast, scalable implementation of the spectrally-accurate boundary integral method developed in \cite{corona2018boundary} was used to simulate active matter systems of up to $\mathcal{O}(10^5)$ particles. This method combines spectral analysis in spherical harmonics bases to perform fast singular and near-singular evaluation of Laplace and Stokes potentials, and fast multipole methods to compute long-range interactions. 

Our contributions in this work to the study of Janus particles through boundary integral simulations are essentially three-fold. First, we extend the method in \cite{corona2018boundary} for efficient scalar potential evaluation to encompass screened Laplace potentials. Second, we couple our scalar evaluation with our rigid body Stokes solver \cite{corona2017integral,yan2019scalable} to explore three dimensional interactions. Finally, we develop well-conditioned boundary integral formulations for three different types of Janus particles that appear frequently in research literature, namely, \emph{amphiphilic}, \emph{bipolar} and \emph{phoretic} particles. These models have a multitude of biomedical applications \cite{su2019janus}. Amphiphilic particles may be employed to model bilipid membranes in cells and to interact with hydrophobic drug particles. Bipolar and phoretic particles are of interest because their motion can be manipulated, via electromagnetic fields and chemical gradients, respectively \cite{rosenthal2012micelle,katuri2017designing}. For each of these test cases, our fast rigid body solver allows us to simulate collective behavior and self-assembly even in densely packed suspensions. 


This paper is organized as follows. In Section \ref{sec:background} we discuss the scalar potentials, introducing notation and relevant boundary integral operators. In Section \ref{sec:discretization}, we develop fast, spectrally accurate methods for boundary integral operator (BIO) evaluation. We then discuss the three classes of Janus particles in detail in sections \ref{sec:amphiphillic} through \ref{sec:Phoretic}, demonstrating how the general set of tools we have developed is adapted to each specific problem.

%% file: background.tex
In this section, we provide the necessary mathematical background and notation for the coupled system of boundary integral equations for long-range Janus and hydrodynamic particle interactions. We present our analysis of the spectra and corresponding evaluation formulas for screened Laplace boundary integral operators (BIO)s employed in the evaluation schemes in Section \ref{sec:Discretization}.

\subsection{Notation}
\label{ssc:notation}
We outline the notation used throughout the paper. In each problem formulation we will consider a system of $M$ rigid spherical Janus particles, which we index with $i=1,...,M.$ For each particle $i$ we adopt the following notation:
\vskip 10 pt
\begin{center}
\begin{tabular}{|c|c|c|c|}

\hline
particle \# & interior & surface & radius \\
\hline
$i$     & $\Omega_i$ & $\Gamma_i$ & $r_i$ \\
\hline
net force & net torque & trans. velocity & rot. velocity \\ 
\hline 
$\boldsymbol{F_i}$ & $\boldsymbol{T_i}$ & $\boldsymbol{v_i}$ & $\boldsymbol{\omega_i}$ \\
\hline
\end{tabular}
\end{center}
\vskip 10 pt

\noindent We will refer to the union of particle interiors and surfaces as $\Omega = \cup_{i=1}^M\Omega_i$ and $\Gamma = \cup_{i=1}^M\Gamma_i$, respectively. The domain exterior to these particles will be denoted by $\Omega_\infty$. For simulations confined in a spherical shell, we denote the surface of the shell by $\Gamma_\8.$ For unbounded simulations, we have that $\Gamma_\8=\emptyset.$

\subsection{Problem setup}
In this work, we will employ an indirect representation of Janus interaction potentials in terms of an unknown integral density defined on particle and geometry boundaries. This Janus potential, which we will denote by $\phi$, satisfies the screened Laplace equation,

\begin{subequations}
\begin{equation}
\label{eqn:SreenedLap}
\nabla^2\phi(\boldsymbol{x}) - \lambda^2\phi(\boldsymbol{x}) = 0 \ \ \forall \, \boldsymbol{x} \in \Omega_\infty.
\end{equation}

This equation models long-ranged interactions which are damped by the medium. The strength of the damping is controlled by the parameter $\lambda$. The quantity $\frac{1}{\lambda}$ has units of length and is referred to as the \textit{Debye Length} in many applications. The standard Laplace equation occurs in models where damping by the fluid medium is negligible, corresponding to $\lambda = 0$.

The boundary condition at particle surfaces is application-dependent; we will show instances of Dirichlet, Neumann, and mixed conditions. For all cases in which $\Omega_\infty$ is unbounded, the potential must also satisfy the decay condition,

\begin{equation}
\label{eqn:freespace}
\lim_{\|\boldsymbol{x}\|\rightarrow \8} \phi(\boldsymbol{x}) = 0.
\end{equation}
\end{subequations}

\paragraph{Janus and hydrodynamic interaction coupling.} For all particle systems considered in this work, the coupling between Janus and hydrodynamic particle interactions involves force and torque balance at particle surfaces. In the phoretic model in Section \ref{sec:Phoretic}, a tangential slip velocity is also induced by chemical activity. These computations involve the evaluation of the Janus force field $-\nabla\phi$. In each application, the associated stress tensor $\mathcal{T}$ can be derived. In electrostatic applications, for instance, $\mathcal{T}$ is the Maxwell stress $\mathcal{T}_{Max}$, given in equation \eqref{eq:Mstress}, with $\epsilon_0$ denoting vacuum permittivity.

\begin{equation}
\label{eq:Mstress}
    \mathcal{T}_{Max}=\epsilon_0(\nabla\phi\otimes\nabla\phi -\frac{1}{2}\norm{\nabla\phi}^2I).
\end{equation}

\noindent 
The net force and torque on particle $i$ are then given by

\begin{equation}\label{eq:Mforcesandtorques}
    \int_{\Gamma_i} (\mathcal{T}\cdot \boldsymbol{\nu}) dS_y = \boldsymbol{F_i}, \quad \int_{\Gamma_i}(\boldsymbol{x}-\boldsymbol{x}_i^c)\times(\mathcal{T}\cdot \boldsymbol{\nu}) dS_y = \boldsymbol{T_i},
\end{equation}

\noindent
where $\nu$ is the outward normal vector to the surface $\Gamma_i$.

\paragraph{Rigid body Stokes problem.} In order to evolve our particulate system, we need to find the rigid body velocities $(\boldsymbol{v}_i,\boldsymbol{\omega}_i)$ which correspond to inter-particle interactions for a given configuration. For a large array of problems, such as the amphiphilic and bipolar particle formulations in Sections \ref{sec:Amphiphilic} and \ref{sec:Bipolar}, the coupling between Janus interactions and hydrodynamic interactions requires force and torque balance at particle boundaries \eqref{eq:Mforcesandtorques}. 

When net forces and torques $(\boldsymbol{F}_i, \boldsymbol{T}_i)$ on rigid bodies are known, the problem of solving the Stokes equation to obtain unknown rigid body velocities $(\boldsymbol{v}_i,\boldsymbol{\omega}_i)$ is known as the rigid body Stokes mobility problem, defined by the system of equations in \ref{eq:Stokes}. In this equation, $\mbu$ denotes the Stokes velocity field and $p$ the corresponding pressure.  We employ the second-kind boundary integral formulation in \cite{corona2017integral,corona2018boundary} to solve this problem. 

\begin{align}
\centering
\label{eq:Stokes}
\begin{split}
-\nabla p + \nabla^2 \mbu &= 0 \quad\text{in}\quad \Omega_\8, \\
\nabla \cdot \mbu &= 0  \quad\text{in}\quad \Omega_\8,  \\
\lim_{\|\boldsymbol{x}\|\rightarrow\infty}\mbu &= 0,    \\
  \mbu &= \boldsymbol{v}^k + \boldsymbol{\omega}^k \times (\mbx - \mathbf{c}^k) \quad\text{on}\quad \Gamma_k, \quad k = 1, \ldots, M,    \\
\int_{\Gamma_k} \mbf \, d\Gamma_k &= -\boldsymbol{F}_k \quad\text{and}\quad \int_{\Gamma_k} (\mbx - \boldsymbol{c}^k) \times \mbf \, d\Gamma_k = -\boldsymbol{T}_k,  \quad k = 1, \ldots, M.
\end{split}
\end{align} 

We note that our rigid body Stokes computational framework is not specific to the mobility problem, and may thus be adapted to other standard rigid body Stokes problems such as the resistance problem, where $(\boldsymbol{v_i},\boldsymbol{\omega_i})$ are known and $(\boldsymbol{F_i}, \boldsymbol{T_i})$ are unknown. It may also be adapted to more general prescriptions at the boundary, as is the case for the phoretic Janus particle model detailed in Section \ref{sec:Phoretic}. 

\subsection{ Boundary integral formulation}
\label{sub:BIEform}
We will employ an integral representation of both Janus potential $\phi$ and velocity field $\mbu$ as a combination of appropriate layer potential boundary integral operators. By design, these representations satisfy the respective PDE and growth conditions. Imposing boundary conditions then yields integral equations for unknown integral densities defined at particle boundaries $\Gamma$ and geometry boundary $\Gamma_\8$.  

\paragraph{Screened Laplace layer potentials.}

For a given value of $\lambda$, let $G_\lambda(\boldsymbol{x},\boldsymbol{y})=\frac{1}{4\pi}\frac{e^{-\lambda\norm{\boldsymbol{x-y}}}}{\norm{\boldsymbol{x-y}}}$ denote the Green's function for the screened Laplace equation. The single and double layer potential operators are defined as follows
\begin{subequations}
\begin{align}
    \SL_\lambda[\sigma](\boldsymbol{x}) &=\frac{1}{4\pi}\int_\Gamma\frac{e^{-\lambda\norm{\boldsymbol{x-y}}}}{\norm{\boldsymbol{x-y}}}\sigma(\boldsymbol{y})d\Gamma, \\
    \DL_\lambda[\mu](\boldsymbol{x}) &=\int_\Gamma \frac{e^{-\lambda \norm{\boldsymbol{x-y}}}\boldsymbol{v_y}^T(\boldsymbol{x-y})}{4\pi \norm{\boldsymbol{x-y}}}\left( \frac{\lambda}{\norm{\boldsymbol{x-y}}}+\frac{1}{\norm{\boldsymbol{x-y}}^2} \right) \mu(\boldsymbol{y})d\Gamma(\boldsymbol{y}). 
\end{align}
\end{subequations}

\noindent Formulas for operators $\SL_\lambda'$ and $\DL_\lambda'$ are similar and can be found in Appendix~\ref{appendix:Derivatives}. It is readily seen that for any $\sigma\in L^2(\Gamma),$ $\SL_\lambda[\sigma](\boldsymbol{x})$ and $\DL_\lambda[\mu](\boldsymbol{x})$ are smooth for $\boldsymbol{x} \notin \Gamma$ and satisfy equation \eqref{eqn:SreenedLap} and condition \eqref{eqn:freespace}. In order to solve a given boundary value problem for $\phi$, we need only find integral densities matching boundary conditions at $\Gamma$. This  motivates a common technique in boundary integral methods \cite{brebbia2012boundary}, in which a potential function $\phi$ is written as a combination of these layer potentials. For example, consider the exterior Dirichlet boundary problem

\begin{equation}
    \nabla^2 \phi(\boldsymbol{x}) -\lambda^2\phi(\boldsymbol{x}) = 0 \ \ \forall\, \boldsymbol{x} \in \Omega_\infty, \ \ \quad \phi=g\; \text{on}\; \Gamma.
\end{equation}

\noindent We propose the ansatz solution $\phi(\boldsymbol{x}) = (\SL_\lambda+\DL_\lambda)[\mu](\boldsymbol{x})$. Taking the limit as $x \rightarrow \Gamma$ in the normal direction, we then use so-called jump conditions for these potentials to obtain a boundary integral equation (BIE) \cite{kress1989linear}:

\begin{equation}
\label{eq:DirichletBd}
    \frac{1}{2}\mu(\boldsymbol{x}) + (\SL_\lambda+\DL_\lambda)[\mu](\boldsymbol{x}) =  g(\boldsymbol{x}) \quad \forall \, x\in\Gamma.
\end{equation}

Careful work is needed to ensure that this ansatz solution yields a uniquely solvable Fredholm equation of the second kind, as is the case for this formulation. This equation is generally well-conditioned, which is highly advantageous for its efficient numerical solution. An accurate and efficient method for solving a discretized version of this equation is presented in Section \ref{sec:Discretization}. 

\paragraph{Integral operators for Stokes problems.} For the Stokes equations, the Stokeslet and Stresslet fundamental solutions are given by: 

\begin{subequations}
\begin{gather}
    G(\boldsymbol{x},\boldsymbol{y})=\frac{1}{8\pi}\left(\frac{I}{\norm{\boldsymbol{x-y}}}+\frac{(\boldsymbol{x-y})\otimes(\boldsymbol{x-y})}{\norm{\boldsymbol{x}}^3}\right),\\
    T(\boldsymbol{x},\boldsymbol{y})=-\frac{3}{4\pi}\left(\frac{(\boldsymbol{x-y})\otimes (\boldsymbol{x-y})\otimes (\boldsymbol{x-y})}{\norm{\boldsymbol{x-y}}^5}\right).
\end{gather}
\end{subequations}
The Stokes single layer potential, its associated traction kernel and the Stokes double layer potential are given by
\begin{subequations}
\begin{equation}
\begin{aligned}
    \SLStk[\sigma](\boldsymbol{x}) &= \int_\Gamma G(\boldsymbol{x},\boldsymbol{y})\sigma(\boldsymbol{y}) d\Gamma(\boldsymbol{y}),\\
    \TStk[\sigma](\boldsymbol{x}) &= \int_\Gamma T(\boldsymbol{x},\boldsymbol{y}) \nu(\boldsymbol{x}) \sigma(\boldsymbol{y}) d\Gamma(\boldsymbol{y}), \\
    \DLStk[\sigma](\boldsymbol{x}) &= \int_\Gamma T(\boldsymbol{x},\boldsymbol{y}) \nu(\boldsymbol{y}) \sigma(\boldsymbol{y}) d\Gamma(\boldsymbol{y}).
\end{aligned}
\end{equation}
\end{subequations}

A number of integral representations have been introduced for rigid body Stokes problems. In this context, we favor representations leading to well-conditioned integral equations (Fredholm of the 2nd kind) and prefer to avoid the introduction of additional unknowns or constraints. We employ the formulation in \cite{corona2017integral} for the Stokes mobility problem, which addresses both issues by representing the flow as the sum of two distinct single layer potentials enforcing force and torque balance, and rigid body motion at particle boundaries, respectively. In Section \ref{sec:Phoretic}, we describe a formulation based on a standard double-layer representation \cite{power1987second} to prescribe a tangential slip velocity at particle boundaries.  

\subsection{Spherical harmonic analysis}
\label{ssc:SHA}
Working with the BIOs for screened Laplace and Stokes in Section \ref{sub:BIEform} requires us to evaluate \emph{weakly singular} and \emph{hyper singular} integrals for targets $x \in \Gamma$. These operators are smooth when evaluated away from the surface, but they become \emph{near-singular}  as a target point $x$ approaches $\Gamma$.  Smooth numerical integration techniques will degrade in quality unless discretization is greatly refined. In \cite{corona2018boundary}, analysis of integro-differential operators in spherical harmonic bases presented in \cite{veerapaneni2011fast,vico2014boundary} was extended to all the Stokes BIOs, and applied to obtain an efficient, spectrally-accurate evaluation scheme for both singular and near-singular cases. We present analysis of BIO signatures and derive evaluation formulas in solid harmonics for the screened Laplace operator. This allows us to extend this fast algorithm framework to the simulation of Janus particles in Section \ref{sec:discretization}. 

\paragraph{Spherical harmonics.}
The \emph{spherical harmonic} of degree $n$ and order $m,$ denoted by $Y_n^m,$ is given by
\begin{equation}
Y_n^m(\theta,\phi)=\sqrt{\frac{2n+1}{4\pi}}\sqrt{\frac{(n-|m|)!}{(n+|m|)!}}P_n^{|m|}(cos\theta )e^{im\phi},
\end{equation}
where $P_n^m$ is the associated Legendre polynomial of corresponding degree and order. The spherical harmonics are eigenfunctions of the screened Laplace equation on the unit sphere, forming an orthonormal basis for $L^2(S^2)$. It follows from a separation of variables argument that any solution to these equations on the interior or exterior of the unit sphere may be written as an expansion of solid harmonics 

\begin{subequations}
\begin{equation}
 \phi_n^m (r, \theta, \phi) = f_n (r) Y_n ^m(\theta, \phi), 
 \label{eq:phi_nm}
\end{equation}
with
\begin{equation}
 \phi(r,\theta,\phi)=\sum_{n=0}^{\infty}\sum_{m=-n}^n \alpha_n^m \phi_n^m(r,\theta,\phi).
 \label{eq:phi_shexp}
\end{equation}
\end{subequations}

\paragraph{Layer potential spectra and evaluation formulas.}
Using the fact that solutions to the screened Laplace equation can be expanded as a superposition of solid harmonics $\phi_n^m$, it can be shown that the spherical harmonics are eigenvectors of $\SL_\lambda$ and $\DL_\lambda$ on the sphere. We present here the eigenvalues for the modified Laplace equation. See Appendix \ref{appendix:Spectra} for a derivation using an argument analogous to that presented in \cite{vico2014boundary}.

\begin{lemma}\label{Spectra}
\label{lemma:spectra}
(Screened Laplace operator spectra).  On the unit sphere, the screened Laplace single and double layer operators diagonalize in the spherical harmonics basis $Y_n ^m$ and their spectra are given by

\begin{center}
$\begin{array} { c | c c c}
&\SL_\lambda & \DL^+_\lambda & \DL^-_\lambda \\[1mm]
\hline \\[1mm]
Y_n^m & \dfrac{ 2\lambda k_n(\lambda) i_n(\lambda )}{\pi} & \dfrac{ 2\lambda^2 i'_n(\lambda) k_n(\lambda) }{\pi} &  \dfrac{ 2\lambda^2 k'_n(\lambda) i_n(\lambda )}{\pi}
\end{array}.$
\end{center}
\end{lemma}

\noindent with $(i_n,k_n)$ the modified spherical Bessel functions of first and second kind, respectively. We use Lemma \ref{lemma:spectra} to evaluate the single and double layer potentials off the surface of the spheres, arriving at the following results. 
\begin{theorem}
\label{thm:LayerSpec}
(Screened Laplace operator evaluation). The single and double layer potentials for density $Y_n^m$ at an arbitrary point off the sphere with spherical coordinates $(r,\theta,\phi)$ are: \\ 

\end{theorem}
\begin{center}
$\begin{array}{c|c|c}
&\SL_\lambda [Y_n^m] (r,\theta,\phi) & \DL_\lambda [Y_n^m] (r,\theta,\phi) \\[3mm]
\hline 
r > 1 & \dfrac{2\lambda i_n(\lambda)}{\pi} k_n(\lambda r) Y_n ^m (\theta,\phi) & \dfrac{2\lambda^2 i'_n(\lambda)}{\pi} k_n(\lambda r) Y_n ^m (\theta,\phi) \\[3mm]
\hline 
r < 1 & \dfrac{2\lambda k_n(\lambda)}{\pi} i_n(\lambda r) Y_n ^m (\theta,\phi) & \dfrac{2\lambda^2 k'_n(\lambda)}{\pi} i_n(\lambda r) Y_n ^m (\theta,\phi) \\[3mm]
\end{array}.$
\end{center}
Given a set of spherical harmonics coefficients $\mu_n^m$ for a given density $\mu$, Theorem \ref{thm:LayerSpec} allows us to evaluate layer potentials on and off the surface. It can be verified that the spectra of $\SL_\lambda'$ and $\DL_\lambda'$ are the derivatives of the above equations with respect to $r$. Formulas are given in Appendix \ref{appendix:Derivatives}. 
The equations of Theorem \ref{thm:LayerSpec} can easily be modified for potentials defined on spheres of different radii. See Appendix \ref{appendix:scaling} for details.

%% file: Discretization.tex
\label{sec:discretization}

As described in the previous section, we rely on representations for $\phi$ and $\mbu$ as a combination of boundary integral operators. Following the work contributed for Stokes boundary integral operators in \cite{corona2018boundary}, we present a spectrally accurate evaluation scheme for the screened Laplace operators. Our goal is to develop a method of applying discretized operators $\SL_\lambda$ and $\DL_\lambda$ efficiently, so that equations like \eqref{eq:DirichletBd} may be efficiently solved by a Krylov subspace iterative method such as GMRES.

\subsection{Janus interaction potential evaluation}
\label{ssc:SolvingJanus}
Any numerical scheme for the approximate solution of BIEs in equation \eqref{eq:DirichletBd} will require accurate evaluation of integrals of the form

\begin{equation}
\boldsymbol{F_i}(\boldsymbol{x}) = \int_{\Gamma_i} K_i(\boldsymbol{x},\boldsymbol{y}) \sigma_i(\boldsymbol{y}) dS(\boldsymbol{y})
\end{equation} 

\noindent for points on and off the surface $\Gamma_i$. The operator of interest will then be the sum over all particles $\boldsymbol{F}(\boldsymbol{x}) = \sum_{i=1}^M \boldsymbol{F_i}(\boldsymbol{x})$. The integral kernel $K_i$ contains a singularity when $x\in\Gamma_i$. To accurately compute $\boldsymbol{F}(\boldsymbol{x}),$ we must be able to evaluate $\boldsymbol{F_i}$ in three regimes: when target points $x$ are far from $\Gamma_i$ (\emph{smooth}), when they are on $\Gamma_i$ (\emph{singular}) and when they are close to $\Gamma_i$ (\emph{near-singular}).

  We split the evaluation of integral densities at each particle into so-called {\em near} and {\em far} fields of targets. For targets in the far field, we employ a standard spectrally convergent smooth quadrature; for large numbers of particles, this computation is accelerated using the Fast Multipole Method (FMM). In the near field, we use the expansions in Theorem \ref{thm:LayerSpec} to evaluate the BIOs of interest.

\paragraph{Smooth integration.} Given a spherical harmonic order $p$, we sample at points $y_{j,k} = y(\theta_j,\phi_k)$ on each sphere with

\begin{equation}
\{ \theta_j = \cos^{-1} (t_j) \}_{j=0}^{p} \ , \ \left\{ \phi_k = \frac{2 \pi}{2p+2} \right\}_{k=0}^{2p+1},
\end{equation}

\noindent with $t_j$ the $(p+1)$ Gauss-Legendre nodes in $[-1,1]$ (\cite{gimbutas2013fast}), for a total of $O(p^2)$ discretization points $y_{j,k}$ per particle, and $N = O(M p^2)$ overall degrees of freedom. To illustrate, let us consider the single layer potential from a single source sphere with surface $\Gamma$ and radius $r$. We have 
\begin{equation}
\SL[\mu](\boldsymbol{x}) =\int_0^{2\pi}\int_0^\pi G(\boldsymbol{x},\boldsymbol{y}(\theta,\phi)) \mu(\theta,\phi) r_i^2 \sin\theta d\theta d\phi
\label{eq:SLSC}
\end{equation}

\noindent for a target point $x$. If the integrand is smooth, this rule will be spectrally convergent as p increases. Although this is the case for any integrand with $x \notin \Gamma_i$, as $x$ approaches the surface, our ability to represent this function and integrate it accurately with order $p$ spherical harmonics degrades. Thus, we only employ this quadrature for targets that are \emph{well-separated} from $\Gamma_i$, that is, such that 

\begin{equation*} 
\mathrm{dist}(\boldsymbol{x},\Gamma) \geq \eta \mathrm{diam}(\Gamma) = 2\eta r_i,
\end{equation*}  
 
\noindent where $\eta$ is determined by a user-defined target accuracy. We will refer to the set of well-separated points from $\Gamma$ as its \emph{far field}, and the complementary set as its \emph{near field}.  


The cost directly evaluating all far field interactions between $M$ particles  is $(p^4 M^2)$. Since this operation is a summation of Green's functions, Fast Multipole Method (FMM) \cite{greengard1987fast} acceleration can be employed to reduce the cost to $O(p^2M)$. The FMM was originally developed for computations involving the Laplace kernel; it has since been extended to many PDEs of interest, including several implementations of the FMM for the screened Laplace kernel \cite{greengard2002new}, as well as for the Stokes kernels \cite{StokesFMM}. We are currently employing the StokesLib3D package \cite{gimbutas2012stfmmlib3} for Stokes interactions and the FMMLIB3D package for Laplace interactions \cite{greengard2012fmmlib3d}. 

\paragraph{Singular integration.} Gauss-Legendre quadrature will fail entirely when target points belong to the source sphere $\Gamma_i$. Such computations are required when calculating particle \emph{self-interactions}. Singular integration on particles of spherical topology may be handled as in \cite{gimbutas2013fast}, using fast spherical grid rotations and FFT acceleration techniques. For spherical particles, the fact that BIOs of interest diagonalize in the spherical harmonics basis allows us to sidestep discretization of singular and hypersingular integrals.

To achieve this, we turn to the spectra of the single and double layer operators, described in Section \ref{ssc:SHA}. Suppose we have a function $\mu$ defined on the unit sphere and we wish to compute $\SL_\lambda[\mu]$. We compute the spectra $\{ \widehat{\mu}_n^m \}$ with a fast forward spherical harmonic transform (SHT). Truncating the spherical harmonic expansion at order $p$, we obtain the approximate equation

\begin{equation}
\label{eq:truncatedsph}
        \SL_\lambda [\mu](\theta,\phi) \approx \sum_{n=0}^p\sum_{m=0}^n \alpha_n^m \widehat{\mu}_n^m Y_n^m(\theta,\phi),
    \end{equation}

\noindent where eigenvalues $\alpha_n^m$ are given by Theorem \ref{thm:LayerSpec}. We evaluate these at all points of the spherical grid via an inverse SHT; FFT-accelerated SHT transforms are applied in $O(p^3\log p)$ operations \cite{mohlenkamp1999fast}. Fast $O(p^2 \log^2 p)$ SHTs are available; however, break-even points are typically large. 

\paragraph{Near-singular integration.} Finally, we must address the evaluation of a potential such as $\phi = \SL_\lambda[\mu]$ at a nearby target point $\boldsymbol{x}$ off the unit sphere with spherical coordinates $(r,\theta,\phi)$. Evaluation formulas in \eqref{thm:LayerSpec} allow us to represent points off the surface of a sphere in terms of spherical harmonics, yielding an expression of the form

\begin{equation}
\phi(\boldsymbol{x}) \simeq \sum_{n=0}^p\sum_{m=0}^n \alpha_n^m \kappa_n^m(r) Y_n^m(\theta,\phi), \label{eq:neval_example}
\end{equation}

\noindent where the vector of coefficients $\boldsymbol{\kappa}$ may be written as $\boldsymbol{\kappa}(r) = \mathcal{F}_\lambda(r) \widehat{\boldsymbol{\mu}}$. Operator $\mathcal{F}_\lambda$ is diagonal with entries dependent only on $r$. The associated flux $\nd{\phi}$ may be evaluated using this same scheme. However, the linear operator mapping $\widehat{\boldsymbol{\mu}}$ to the spherical harmonic coefficients  will be tridiagonal and its entries will depend on $\theta$ and $\phi$ as well as $r$. 

We use this scheme to evaluate interactions between a particle and target points intersecting its near field. Parameters $\eta$ and $p$ must be chosen carefully to balance accuracy and cost of near and far evaluation routines. Since the number of particles neighboring a fixed particle is bounded, the maximum number of near field target points per particle is $O(p^2)$. Direct evaluation of \eqref{eq:neval_example} is thus $O(p^4)$ per particle. Efficient $O(p^3 \log p)$ accelerations based on FFTs or translation operators are proposed in \cite{corona2018boundary}. 

\subsection{Computation of scalar influence on fluid flow}
\label{ssc:derivatives}
To evaluate the coupling between Janus interactions and the fluid, we need to compute $\nabla\phi$. We decompose the gradient as 
\begin{equation}
\label{eqn:grad_pot}
\nabla\phi = \nabla_\Gamma\phi + \nd{\phi}\boldsymbol{n},
\end{equation}
where $\nabla_\Gamma$ is the surface gradient. $\nabla_\Gamma\phi$ can be expressed in terms of angular derivatives of the first fundamental forms of the surface. For spherical bodies, these values may be computed explicitly. Since the shapes do not change with time, the surface gradient operator remains fixed and may be precomputed as a matrix. To compute the normal derivative, we again employ properties of the single and double layer operators. For instance, for the ansatz proposed in Section \ref{sub:BIEform}, we have that 
\begin{equation}
\nd{\phi} = (\SL'_\lambda + \DL'_\lambda)[\mu]
\end{equation} 
in the exterior of the particles and 
\begin{equation}
\nd{\phi} = (\SL'_\lambda + \DL'_\lambda)[\mu] -\frac{1}{2}\mu
\end{equation}
on the surface of the particles.
These quantities are computed using the same technique described in \ref{ssc:SHA}, with the operator spectra given in Appendix \ref{appendix:Derivatives}. 
 $\nabla\phi$ may thus be computed at the cost of multiplication by a precomputed matrix and one additional layer potential evaluation.    
\subsection{Particle evolution and collision resolution}
\label{ssc:evolve_collision}
As was outlined in Section \ref{sec:background}, the Janus interaction potential $\phi$ is formulated as a combination of layer potentials; given a configuration of $M$ rigid Janus particles, the evaluation routines in Section \ref{ssc:SolvingJanus} allow us to efficiently solve the corresponding BIE via a Krylov subspace method. Given the resulting net forces and torques in \eqref{eq:Mforcesandtorques}, we then employ techniques in \cite{corona2018boundary} to solve the relevant Stokes rigid body problem and find translational and rotational velocities. Finally, standard explicit time discretization schemes are used to evolve rigid particle positions and rotation frames in time. This sequence is repeated each timestep; the resulting particle evolution algorithm is illustrated in Figure \ref{fig:time_cycle}.

When particles are in close proximity to each other or to the geometry, potential collisions must be detected and resolved. This must be done carefully to preserve both cost-efficiency and physical fidelity. Following the approach in \cite{yan2019scalable}, we employ a linear complementary formulation (LCP) of contact; for each colliding pair of particles, a normal force must be applied to prevent interpenetration. In this work, we employ the state-of-the-art Barzilai-Borwein Projected Gradient Descent method (BBPGD) to find the unknown contact forces, adding them to Janus interaction forces and torques for particle evolution \cite{fletcher2005barzilai}. We note that each iteration of the PGD involves solving a Stokes rigid body problem; in our experiments, the number of iterations remains small ($\lesssim 10$ PGD iterations).

\tikzstyle{block} = [rectangle, draw, text width=10em, text centered, rounded      corners, minimum height=3em]
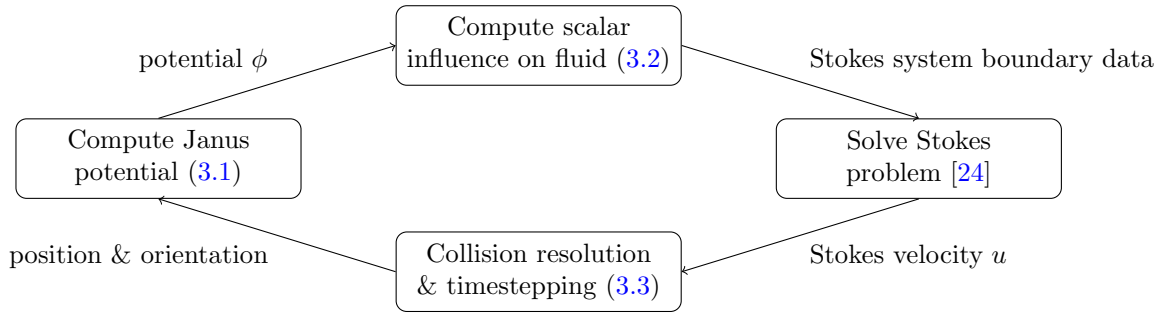
\begin{figure}[h]
\begin{tikzpicture}
\centering
 [node distance=1.35cm,
 start chain=going below,]
\node (n1) at (0,1.5) [block]  {Compute Janus potential (\ref{ssc:SolvingJanus})};
\node (n2) at (5,0) [block] {Collision resolution \& timestepping (\ref{ssc:evolve_collision})};
\node (n3) at (10,1.5) [block] {Solve Stokes problem \cite{corona2017integral}};
\node (n4) at (5,3) [block] {Compute scalar influence on fluid (\ref{ssc:derivatives})};


\draw [->] (n1.north) -- (n4.west) node[midway,above left] {potential $\phi$};
\draw [->] (n4.east) -- (n3.north) node[midway,above right] {Stokes system boundary data};
\draw [->] (n3.south) -- (n2.east) node[midway,below right] {Stokes velocity $u$};
\draw [->] (n2.west) -- (n1.south) node[midway,below left] {position \& orientation};
\end{tikzpicture}
\caption{{\em Advancing the system by a single time step requires us to: (i) solve the Janus interaction potential BIE, (ii) setup and solve the corresponding Stokes rigid body problem, (iii) use the Stokes formulation to resolve collisions via the LCP, and (iv) given rigid body velocities, advance particle positions and orientations.} }
\label{fig:time_cycle}
\end{figure}
\subsection{Validation}
\begin{figure}[H]
\centering
\includegraphics[width=\textwidth]{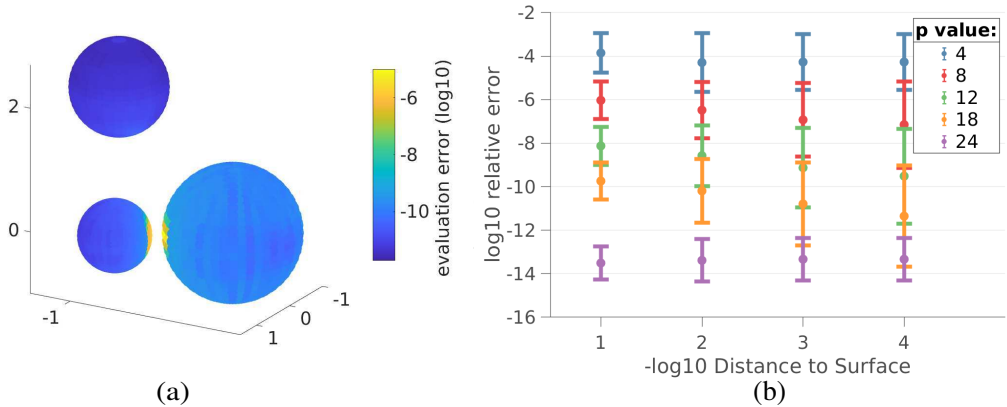}
\caption{ {\em Evaluation test consisting of three spheres with centers  (-1.5,1,0),(-3,0,0),(-1,0 2) and radii 0.5,1,0.7, respectively  (a). We evaluate the potential at spherical target shells located at distance $(10^{-k}) r_i$ from each surface, for $k=1,...,4$. We plot the relative error distribution, and observe that it does not increase as target points approach particle surfaces, and it decays spectrally as $p$ increases (b). }}
\label{fig:BIE_validation}
\end{figure}

We construct a simple boundary value problem to test the accuracy of our near evaluation routine. For a system of 3 spheres of different radii, we consider the potential induced by point charges randomly placed inside each sphere; this potential can be easily evaluated at targets $x$ as a weighted sum of Green's functions $G(\boldsymbol{x},\boldsymbol{x_i})$ for each point charge $\boldsymbol{x_i}$. To test our evaluation scheme, we evaluate the potential induced on each surface, solve the Dirichlet BIE in \eqref{eq:DirichletBd}, and use the integral representation to evaluate it at shells of target points at distance $(10^{-k}) r_i$ from each spherical surface. The results in Figure \ref{fig:BIE_validation} demonstrate spectral accuracy independent of the distance between the surface and target. 

%% file: amphiphillic.tex
\label{sec:amphiphillic}
Amphiphilic particles are split into a hydrophilic head and hydrophobic tail. Suspensions of such particles serve as mimetic models for cell membrane dynamics and are widely used in self-assembling nanomaterials. We first describe an integral formulation for amphiphilic Janus interactions. We then use our simulation framework to demonstrate spontaneous self-assembly of micelles in three dimensional systems.

\subsection{Formulation}

We employ the model developed by Fu et. al \cite{fu2019amphiphilic}, which formulates hydrophobic interaction potentials in integral form. The hydrophobic interaction potential $\phi$ is defined as the smooth minimizer of the hydrophobic energy functional for a given particle configuration and Dirichlet boundary value $f$:
\

\begin{equation}
\label{eq:Ifunc} 
E[\phi]=\int_{\Omega_\8} |\nabla\phi|^2 + \frac{1}{\rho^2} \phi^2 dV = \int_\Gamma \phi\frac{\partial\phi}{\partial\nu}d\Gamma.  
\end{equation}
The equality of the two integrals follows from Green's identities and equation \eqref{eq:hydrophilic_dirichlet} below.   
$E$ is derived from a quadratic expansion of the film tension on a sphere, the details of which are given in sources such as \cite{marvcelja1977role,eriksson1989phenomenological}; it is used to investigate lipid membrane interactions in \cite{ryham2016calculating}. Through variational methods, $E$ can be shown to have a unique minimum which satisfies the screened Laplace equation 
    \begin{equation}
    \label{eq:hydrophilic_dirichlet}
        -\rho^2\nabla^2\phi(\boldsymbol{x}) + \phi(\boldsymbol{x})=0 \quad \boldsymbol{x}\in \Omega_\8 , \quad \phi(\boldsymbol{x})=f \quad \boldsymbol{x}\in \Gamma,
    \end{equation}
    
\begin{figure}[H]
    \centering
    \includegraphics[width = .75\textwidth]{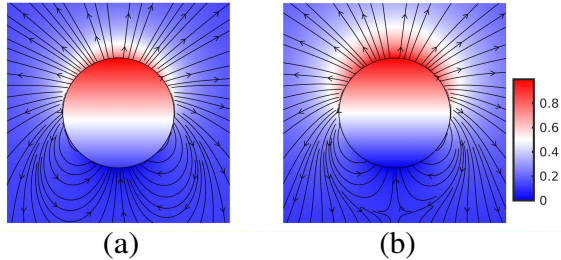}
    \caption{{\em Amphiphilic potential of a particle with $\lambda = 1$ (a) and $0.1$ (b) with field lines drawn. The comparison between these figures illustrates the effect of $\lambda$ on the potential decay due to the conductive properties of the fluid medium}}
    \label{fig:single_particle_lambda}
\end{figure}

\noindent where $f$ describes the hydrophobic character of the boundary with $0<f<1$ and $\rho$ is a characteristic length of attraction. We recover the screened Laplace equation in standard form by setting $\lambda =\frac{1}{\rho}$.  In our simulations we take $f$ to be a shifted cosine function, $f(\theta)=\frac{1}{2}(\cos{\theta} + 1)$, where $\theta$ is the co-latitude of a point on the particle surface.  Figure \ref{fig:single_particle_lambda} shows a cross-section of the resulting potential for two values of lambda.  

\paragraph{Integral equation formulation.} As the hydrophobic potential $\phi$ satisfies the screened Laplace equation, we make use of the ansatz proposed in equation \eqref{eq:DirichletBd}; the exterior Dirichlet problem in equation \eqref{eq:hydrophilic_dirichlet} leads to the BIE

\begin{equation}
 f(\boldsymbol{x}) = \frac{1}{2}\mu(\boldsymbol{x}) + (\SL_\lambda +\DL_\lambda)[\mu](\boldsymbol{x})\quad \boldsymbol{x}\in\Gamma.
\end{equation}

\noindent The corresponding stress tensor \cite{fu2019amphiphilic} can be shown to be

\begin{equation}
\label{eqn:Amphi_Tensor}
    \mathcal{T}_J =\eta\left(\frac{\phi^2}{\rho^2}I+2\left(\frac{1}{2}|\nabla \phi|^2-\nabla \phi \otimes\nabla \phi\right)\right).
\end{equation}

In this expression, all of the units are nondimensionalized and $\eta$ is the ratio of the amphiphilic to viscous pressure,  $\eta = \pi_a / \pi_{v}$. Expressions for these pressures are presented in appendix \ref{app:Nondim}.

\subsection{Self Assembly}

Employing our boundary integral approach, we simulate self-assembly in systems of amphiphilic particles. In all of these examples we use the average particle radius, $ r$, as the characteristic unit length. If we let the particle radius be $1 nm$, with  $\gamma = 1 \frac{pN}{nm}$ and $\mu = 1 cPas$ then the time unit is $10^{-8} s$, where nondimensionalized time is given by the expression $t =\frac{r\mu}{\gamma}$. We have chosen $\Delta t = 0.1$ in these simulations, so that one timestep corresponds to $ 1\times 10^{-9}$ seconds. Likewise, we report nondimensionalized energy so that one unit corresponds to $\gamma$ times unit length squared. In the following examples this corresponds to $10^{-21}$ joules. These parameter values are used in the following experiments, except where otherwise stated.

\begin{figure}[H]
    \centering
    \includegraphics[width = \linewidth]{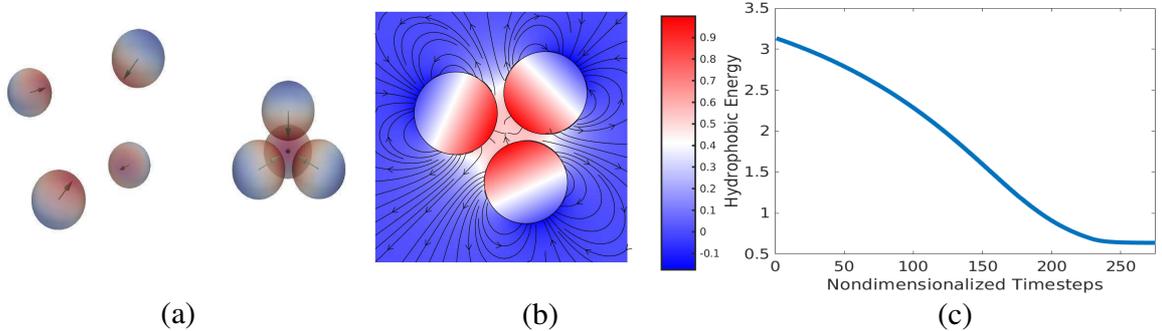}
    \caption{{\em Systems of four amphiphilic particles form a tetrahedral configuration (a). We plot the hydrophobic potential on a cross section through a base of the tetrahedron (b), illustrating the higher potential region between particles. Finally, we plot the hydrophobic energy as the tetrahedron forms (c)}} 
    \label{fig:tetra_panels}
\end{figure}

\paragraph{Four particle system.}
To illustrate this phenomenon, we first present a small example involving four amphiphilic spheres of the same size; these are initialized with random initial positions and orientations. In such a situation, the particles are known to form a tetrahedral configuration. Figure \ref{fig:tetra_panels} shows the initial and final configurations of the spheres as well as cross sectional plots of the resulting potential, agreeing with two dimensional results in \cite{fu2019amphiphilic,zhang2017janus}. The particles seek to minimize contact between the fluid and the hydrophobic tails by shielding them in the center of the configuration. 

\begin{figure}[H]
    \centering
    \includegraphics[width = \linewidth]{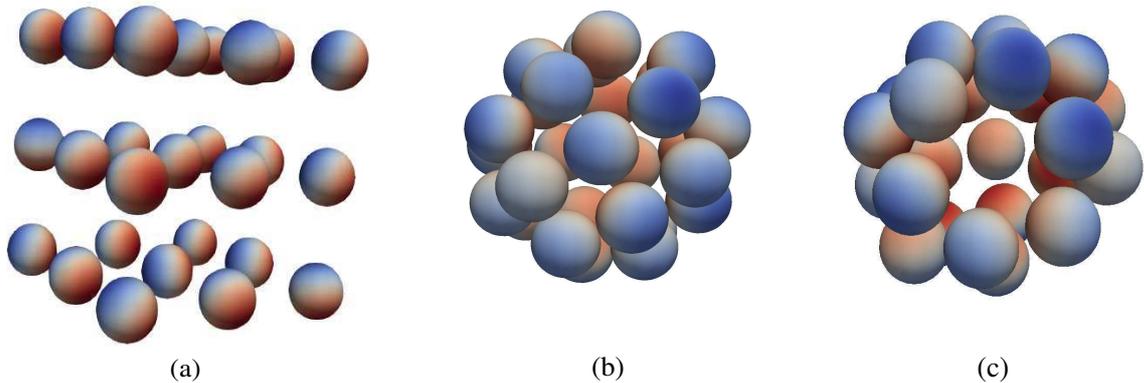}
    \caption{{\em Three dimensional configurations of particles tend to form micelles, with their hydrophobic ends pointing inwards. We demonstrate with a $3\times 3\times 3$ lattice of particles with random initial orientation (a) and the resulting structure (b). When three particles are removed from the configuration, the resulting configuration is less tightly packed (c).}}
    \label{fig:amphiphilic_cube}
\end{figure}

\paragraph{Micelle formation.} 
We then explore the dynamics of larger systems of amphiphilic particles. In \cite{fu2019amphiphilic} the authors studied the long-term configurations resulting from two dimensional amphiphilic interactions. We recreate their experiment here in three dimensions. Figure \ref{fig:amphiphilic_cube} shows the final configurations for two instances of the corresponding three dimensional experiment; in both, a single layer structure (micelle) forms as particles cluster with their hydrophobic ends facing inwards.

We observe micelles to be the most common long-term configurations; their formation does not appear to depend strongly on initial conditions or the number of particles. However, the resulting micelles are more tightly packed for certain numbers of particles. We observe the formation of other stable structures, such as bilayer sheets, when the initial configuration is sufficiently close to the final configuration. In two dimensional experiments, bilayers are observed to form spontaneously \cite{fu2019amphiphilic}. We do not notice such spontaneity across our experiments in three dimensions. 

%% file: Bipolar.tex

We present a model for bipolar electric Janus particles. These particles display concentrations of charge density of opposite signs on their northern and southern hemispheres. Such particles have been shown to exhibit self-assembly behavior such as the formation of chains \cite{pawar2010fabrication}, and can be manipulated through careful application of electric and magnetic fields \cite{velev2009particle,seong2016magnetic}. 

\subsection{Formulation}
\label{ss:BPJP}

We assume particle interiors are perfect conductors and their interactions are electrostatic. We wish to follow the model employed in \cite{hossan2015bipolar} in which a constant electric field $E_0$ is applied. However, if the fluid is an imperfect conductor ($\lambda>0$), a constant electric field is not physical. To resolve this, we confine our experiments to the interior of a rigid shell, $\Omega_{sh} = \Omega \cup \Omega_\infty$ with boundary $\Gamma_{\infty}$, allowing a constant field applied to the shell boundary to permeate into the fluid. In this setting, Maxwell's equations reduce to coupled Laplace (particle interiors) and screened Laplace (exterior) equations for $\phi$, the scalar electrostatic potential. Particle interiors are assumed to have uniform electric permittivity, $\epsilon_i$, while the exterior has permittivity $\epsilon_0$. We normalize the permittivity by dividing through by $\epsilon_0$, so that the exterior permittivity is 1 and the interior permittivity is $\epsilon= \epsilon_i / \epsilon_0$. 

\paragraph{Mathematical model.}
 Gauss's Law states that for a charge distribution $\rho$, the resulting electrostatic force potential $\phi$ must satisfy
\begin{equation}
    \nabla\cdot(\epsilon\nabla\phi) = \rho.
\end{equation}

\noindent In the particle interior, Gauss's law simplifies to a Poisson equation.
 The charge distribution, $\rho,$ is prescribed at the start. We choose $\rho$ to be the charge induced by a pair of point charges of equal strength and opposite sign in the interior of the particle. In the exterior, we model the electrostatic potential described by the linearized Poisson-Boltzmann equation, the derivation of which can be found in \cite{gilson1993computation}. We arrive at the following system of equations with boundary conditions:

\begin{alignat}{2}
\label{eq:Bipolar_Lap}
\nabla^2\phi &= \frac{1}{\epsilon} \rho, \quad \boldsymbol{x}\in \Omega, & \ \ \
\nabla^2\phi - \phi &= 0, \quad \boldsymbol{x}\in\Omega_{\infty} \\
 [[\phi]]_\Gamma &= 0 & \ \ \ 
\left[ \left[ \epsilon(\boldsymbol{x})\nd{\phi} \right] \right]_\Gamma &= 0.\\
\end{alignat}

\noindent All physical quantities have been nondimensionalized in the manner described in Appendix \ref{app:Nondim}.

\paragraph{Boundary integral formulation.}
We derive a novel boundary integral equation formulation for the electrostatic potential $\phi$. A similar direct second kind formulation based on Green's theorems was derived and analyzed in \cite{geng2013treecode}. This system has mixed boundary conditions and requires the evaluation of interior potentials. We represent the potential with a pair of unknown densities, $\psi$ and $\mu$. The potential induced from point charges in the interior is denoted by $\mathcal{Q}.$ The potential from the exterior field, $\mathcal{E},$ is represented as a constant electric field in the exterior of the shell and as a layer potential in the shell interior:  
\begin{alignat}{2}
   \mathcal{E}(\boldsymbol{x}) &= \boldsymbol{E}^T_0\boldsymbol{x} \quad \boldsymbol{x}\in\Gamma_\8, \\
   \mathcal{E}(\boldsymbol{x}) &= (\SL_\lambda + \DL_\lambda)[\mu_\8](\boldsymbol{x}) \quad \boldsymbol{x}\in\Omega_\8.
\end{alignat}
$\mu_\infty$ is determined by equating the two expressions at the boundary and solving the resulting integral equation at the beginning of our simulation. We then make the following ansatz, expressing $\phi$ as

\begin{subequations}
\begin{equation}
\begin{aligned}
\label{eqn:potential_rep}
\phi(\boldsymbol{x}) &=\SL_0 [\psi](\boldsymbol{x}) + \DL_0[\mu](\boldsymbol{x}) + Q(\boldsymbol{x}) \quad \boldsymbol{x}\in\Omega, \\
\phi(\boldsymbol{x}) &=\SL_\lambda[\psi](\boldsymbol{x}) + \DL_\lambda[\mu](\boldsymbol{x}) + \mathcal{E}(\boldsymbol{x}) \quad \boldsymbol{x}\in\Omega_{\infty}.
\end{aligned}
\end{equation}
\end{subequations}

\noindent This formulation automatically satisfies the Poisson and Poisson-Boltzmann equations on the respective domains. Enforcing the jump conditions, we obtain the following system of equations for $\boldsymbol{x} \in \Gamma$: 
\begin{subequations}
\begin{equation}
	\begin{aligned}
    0 &=(\SL_\lambda-\SL_0)[\psi](\boldsymbol{x}) +(\DL_\lambda - \DL_0)[\mu](\boldsymbol{x}) +\mu(\boldsymbol{x}) - (\mathcal{E} - Q)(\boldsymbol{x}), \\
    0 &=(\SL'_\lambda - \SL'_0)[\psi](\boldsymbol{x}) + (\DL'_\lambda - \DL'_0)[\mu](\boldsymbol{x}) -\psi(\boldsymbol{x})/2 -\epsilon\psi(\boldsymbol{x})/2 -  \left(\nd{\mathcal{E}} - \epsilon\left(\nd{Q} \right)\right)(\boldsymbol{x}).
\end{aligned}
\end{equation}
\end{subequations}

\noindent We may then set the following matrix equation by evaluating $Q$ and $\nd{Q}$ on the boundary
\begin{equation}
\begin{pmatrix}
I + \DL_\lambda - \DL_0,\quad & \SL_\lambda - \SL_0 \\
\DL'_\lambda - \epsilon \DL'_0 & -\frac{(1+\epsilon)}{2}I + \SL'_\lambda - \epsilon \SL'_0  \\
\end{pmatrix} 
\begin{bmatrix}
\mu\\
\psi
\end{bmatrix} 
= \begin{pmatrix}
        \mathcal{E} - Q\\
        \epsilon\left(\nd{\mathcal{E}} -\nd{Q}\right)
\end{pmatrix}.
\label{eq:BIEbimetallic}
\end{equation}

This is a coupled, second-kind system of BIEs for densities $\mu, \psi$. Once solved, we use expressions in equations \eqref{eqn:potential_rep} to find the potential $\phi$ at arbitrary points. We use the Maxwell stress tensor to compute forces and torques. Rigid particle translational and rotational velocities are then computed by solving the Stokes mobility problem.

\subsection{Numerical Experiments}
In Figure \ref{fig:diagonal} we reproduce as closely as possible a two dimensional experiment presented in \cite{hossan2015bipolar}. Bipolar particles are initially placed in a diagonal arrangement. In this setup the time nondimensionalization is given by $t=\frac{q_c^2\mu}{\epsilon_0\|\boldsymbol{E_0}\|^2}$. In our experiments we use $cC$ as the base unit of charge and $0.01 V/\mu m$ as the units of field strength, as well as taking the viscosity $\mu$ to be 1 $mPas$. We take the unit length to the radius of a particle, which we set to be $1\mu m.$ The radius of the confined system, $r_{sh}$ is set to 25 units. The charge orientation of each particle is initially aligned with the x axis, perpendicular to the external field, which points in the direction of the negative y axis. In terms of our nondimensionalized units, we have $q_c = 50$ and $\|\boldsymbol{E_0}\|=10$. 

\begin{figure}[H]
\captionsetup[subfigure]{labelformat=empty}
		\centering 
		\subfloat[][]{
		\includegraphics[width=0.75\textwidth]{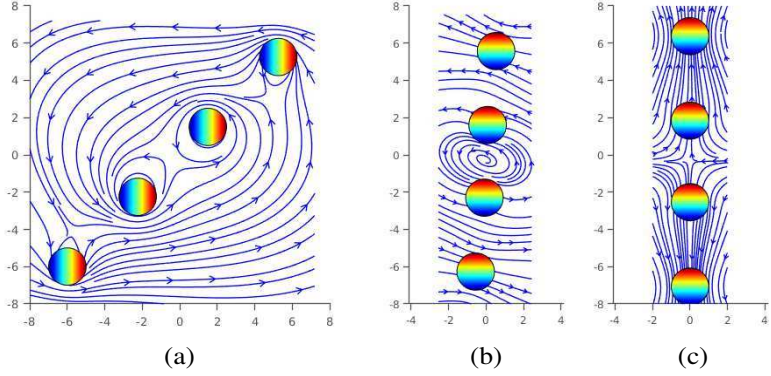}
		\label{fig:align_finaler}}
	\caption{{\em We simulate an experiment described in \cite{hossan2015bipolar}, where bipolar electric particles were placed in the configuration (a). Particles are colored according to surface charge; red and blue represent positive and negative, respectively. A constant electric field is applied in the positive y direction, causing the particles to rotate in the direction of the field (b), while the entire line moves to align with the field (c). Fluid streamlines illustrate the initial rapid rotation of the individual particles and slower rotation of the chain. The maximum fluid speed of the second and third panels are, 3\%, and 0.5\% of the maximum initial speed, respectively.}}
	\label{fig:chain_streamlines}
\end{figure}

The electric force imbalance on the particles induces clockwise rotation of both the particles and the line, as the particles form a chain in the direction of the induced electric field. Figure \ref{fig:chain_streamlines} shows the streamlines from the flow field that result from particle interaction. Initially, rapid local rotations are present in the fluid as the particles rotate. After this occurs, the flow becomes more globally rotational, and the particles form a chain aligned in this direction.

\begin{figure}[H]
	\centering
\includegraphics[width = \linewidth]{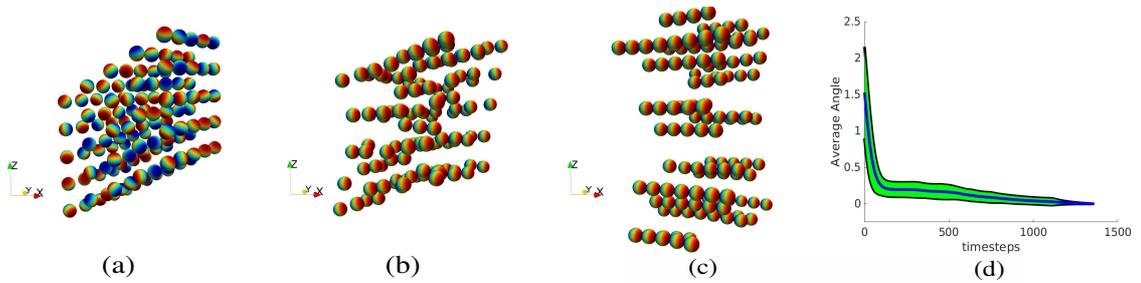}
	\caption{ {\em 125 bipolar particles are oriented randomly (a). When an external electric field is applied in the direction of the x axis, the particles orient themselves in that direction (b) and form chains (c). To quantify polarization, we plot the distribution of angles that the particles make with the electric field (d).}}
	\label{fig:125_chain}
\end{figure}

\paragraph{}

We then conducted larger-scale experiments to study spontaneous chain formation in three dimensions; snapshots of one such experiment are shown in Figure \ref{fig:125_chain}. We place 125 particles with random orientations and with initial positions on a lattice offset from the direction of the electric field. Initially, a locally rotational flow forms as particles rotate to align with the electric field, whereupon chain formation is observed. As chains form, they also begin to repel each other. Throughout this simulation, we quantify the extent of polarization by plotting the distribution of particle angles relative to the background field, as is shown also in Figure \ref{fig:125_chain}d.

Across all our three dimensional experiments, we observe spontaneous chain formation irrespective of initial positions or orientations, closely matching previous results in two dimensional studies. This confirms that, by changing the external field, one can effectively control the orientation of the particles and induce chain formation.

%% file: phoretic.tex
 Phoretic particles are a class of Janus particles with interactions driven by fluid slip on their surface; in this work, we discuss a type of phoretic particle driven by chemical reactions with a solute. Phoretic particle suspensions are useful for modelling microswimmers, particles that propel themselves by ``pushing'' or ``pulling'' the surrounding fluid. Spherical phoretic particles are also of great interest as drug delivery mechanisms in biological systems. 
 
 We employ a standard mathematical formulation for phoretic particles, described in \cite{kanso2019phoretic}. This formulation models phoretic particle interactions via diffusion of chemical concentrations in solution inducing a tangential slip velocity on particle surfaces. As the main coupling between Janus and hydrodynamic interactions is due to this tangential slip, the resulting rigid body problem is not a mobility problem; we detail an integral formulation for the resulting Stokes problem. 

\subsection{Formulation}
In the model we employ, the phoretic character of a particle is determined by two functions on the particle surface, $A(\theta)$ and $M(\theta).$ $A(\theta)$ governs the flux of chemical concentration at each particle surface, while $M(\theta)$ models how concentration gradients induce tangential slip on the fluid. 

Consider $M$ rigid spherical Janus particles suspended in a fluid with viscosity $\mu$ inside a closed domain $\Omega_\infty$. Let $\{\Omega_k, \Gamma_k\}_{k=1}^M$ denote the domains and boundaries of the rigid particles respectively. The chemical concentration $C$ is determined by solving a Laplace Neumann boundary value problem:
\begin{subequations}
\begin{align}
\nabla^2 C &= 0  \quad\text{in}\quad \Omega_\infty , & \\
\frac{dC}{dn} &= A_\infty \quad\text{on} \quad \Gamma_\infty, &   \\
  \frac{dC}{dn} &= -A_k(\theta_k) \quad\text{on}\quad \Gamma_k, \quad k = 1, \ldots, N.    
 \end{align} 
 \label{eq:concentration}
\end{subequations}%
Solutions must satisfy the compatibility condition that $\int_{\Gamma_\infty} A_\infty =\sum_{k=1}^M \int_{\Gamma_k} A_k dS_k = 0$. In an unbounded context, the flux condition on the boundary is replaced with the far field condition that $\lim_{\|x\|\rightarrow\infty}C(x) = 0.$  
The concentration gradient induces a tangential slip velocity, given by 

\begin{equation}
\boldsymbol{u}_{slip}^k = M_{k}(\theta_k)(I-\boldsymbol{n}\boldsymbol{n^T})\nabla C\quad \text{on} \quad \Gamma_k.
\end{equation}
The corresponding equations for the fluid velocity are given by:
\begin{subequations}
\begin{align}
\label{eq:phoretic_system}
-\nabla p + \nabla^2 \mbu &= 0 \quad\text{in}\quad \Omega_\infty, & 
\\
\nabla \cdot \mbu &= 0  \quad\text{in}\quad \Omega_\infty, & \\
  \mbu &= 0 \quad\text{on}\quad \Gamma_\infty, &   \\
  \mbu &= \mbu_{slip}^k + \boldsymbol{v}^k + \boldsymbol{\omega}^k \times (\boldsymbol{x}^k - \boldsymbol{c}^k) \quad\text{on}\quad \Gamma_k, \quad k = 1, \ldots, N,    \\
\int_{\Gamma_k} \boldsymbol{f} \, d\Gamma_k &= -\boldsymbol{F}_k \; \text{and} \; \int_{\Gamma_k} (\boldsymbol{x^k} - \boldsymbol{c}^k) \times \boldsymbol{f} \ d\Gamma_k = -\boldsymbol{T}_k, \quad k = 1, \ldots, N.
\end{align}
\label{eq:governing}
\end{subequations}%

This system closely resembles the formulation of the Stokes mobility problem in equations \eqref{eq:Stokes}. However, in this case, the Laplace potential is mainly coupled to the Stokes equation through an induced slip velocity, rather than through rigid body forces and torques, $\mathbf{F}_k$ and $\mathbf{T}_k$, which are both equal to zero unless particles are in contact with each other or the domain boundary $\Gamma_\8$. We present an integral representation tailored to this tangential slip problem below.

\subsection{Boundary integral formulation}

In this case, the scalar potential $\phi$ corresponding to Janus particle interactions
must satisfy the Laplace Neumann BVP in \eqref{eq:concentration}; we follow the standard approach for this problem representing it as a single layer potential defined on $\Gamma \cup \Gamma_\8$ \cite{kress1989linear}. The Stokes potential in this case is more involved. We outline the steps below.

\paragraph{Stokes integral formulation.} We begin by making the ansatz that the fluid velocity $\boldsymbol{u}(\boldsymbol{x})$ can be expressed as 
\begin{equation}
\label{eq:Phoretic_u}
\boldsymbol{u}(\boldsymbol{x}) = \DLStk_\infty[\boldsymbol{\mu_\infty}](\boldsymbol{x}) + \sum_{k=1}^M (\DLStk_k + \mathcal{V}_k)[\boldsymbol{\mu_k}](\boldsymbol{x}),
\end{equation}

\noindent where $\mathcal{V}_\8 [\mu_\8] $ is a rank 1 correction for the Stokes double layer interior operator given by

\begin{subequations}
\begin{equation}
\mathcal{V}_\infty[\boldsymbol{\mu}](\boldsymbol{x}) = \frac{1}{4\pi} \boldsymbol{e_r}(\boldsymbol{x})\int_{\Gamma_\infty}(\boldsymbol{\mu}\cdot \boldsymbol{e_r}) dS
\end{equation}

\noindent on the bounding surface and $\mathcal{V}_k [\mu_k]$ is a standard completion flow , with

\begin{equation}
\mathcal{V}_k[\boldsymbol{\mu}](\boldsymbol{x}) = G(\boldsymbol{x-c_k})\int_{\Gamma_k}\boldsymbol{\mu}(\boldsymbol{x})dSy + R(\boldsymbol{x-c_k})\int_{\Gamma_k}(\boldsymbol{y-c_k})\times \boldsymbol{\mu}(\boldsymbol{y})dS(\boldsymbol{y})
\end{equation}
on the surface of each particle \cite{power1987second}.
\end{subequations} 

By substituting these expressions into equations \eqref{eq:Phoretic_u} and taking the limit as $x$ approaches each component of the boundary in the normal direction, we obtain a second kind BIE. The force and torque balance boundary condition remain the same, with
\begin{equation}
\label{eq:forcetorqueconstraints}
\int_{\Gamma_k} \boldsymbol{\mu} dS_k = \boldsymbol{F_k},\quad \int_{\Gamma_k} \boldsymbol{\mu} \times (\boldsymbol{x}-\boldsymbol{c^k}) dS_k = \boldsymbol{T}_k. 
\end{equation}
Overall, we have a system of BIEs for the Stokes equations of the form: 
\begin{center}
\begin{equation}
\begin{bmatrix}
-\frac{1}{2} I + \DLStk_\infty + \mathcal{V}_\8 & \sum_{k=1}^M (\DLStk_{k,\infty} + \mathcal{V}_{k,\infty}) & 0 \\
 \DLStk_{\infty,k} & (\frac{1}{2} I + \sum_{k=1}^M(\DLStk_k + \mathcal{V}_k)) & -G \\ 
  0 & H & 0 \\ 
\end{bmatrix} 
\begin{bmatrix} \boldsymbol{\mu}_\infty \\ \boldsymbol{\mu}_{RB} \\ \boldsymbol{V}_{RB} \end{bmatrix}
= 
\begin{bmatrix} 0 \\ \boldsymbol{U}_{slip} \\ \boldsymbol{F}_{RB} 
\end{bmatrix}.
\label{StokesBIEsystem_Phoretic}
\end{equation}
\end{center}
Here, $\boldsymbol{U}_{slip}$ is a vector of the slip velocities on each particle,  $\boldsymbol{V}_{RB}$ is a vector consisting of $\nu$ and $\omega$ for each particle, and $\boldsymbol{F}_{RB}$ is a vector of corresponding rigid body forces and torques.  $G$ is a block-diagonal operator mapping $\boldsymbol{V}_{RB}$ to rigid body motion velocities at particle boundaries and $H$ is a block-diagonal operator that computes the two integrals in \eqref{eq:forcetorqueconstraints}.

The system of equations in  \eqref{StokesBIEsystem_Phoretic} is then solved at every timestep with $\boldsymbol{F}_{RB}$ set to zero. If other forces are present, such as those from contact resolution, \eqref{StokesBIEsystem_Phoretic} is then solved a second time with nonzero $\boldsymbol{F}_{RB}$, using the first solution and complementarity to facilitate this second solve.

\subsection{Results}

A wide range of phoretic particles can be modeled by the functions $A$ and $M$. For our studies, we will focus on simulating systems of so-called ``Saturn particles'' \cite{golestanian2007designing}, which are defined by prescribing 
\begin{equation}
{A}_k(\theta) = a_k (1-\cos^2 \theta), \quad {M}_k(\theta) = m_k \cos \theta.
\label{saturnlabels} 
\end{equation}

\begin{figure}[H]
    \centering
    \includegraphics[width = \linewidth]{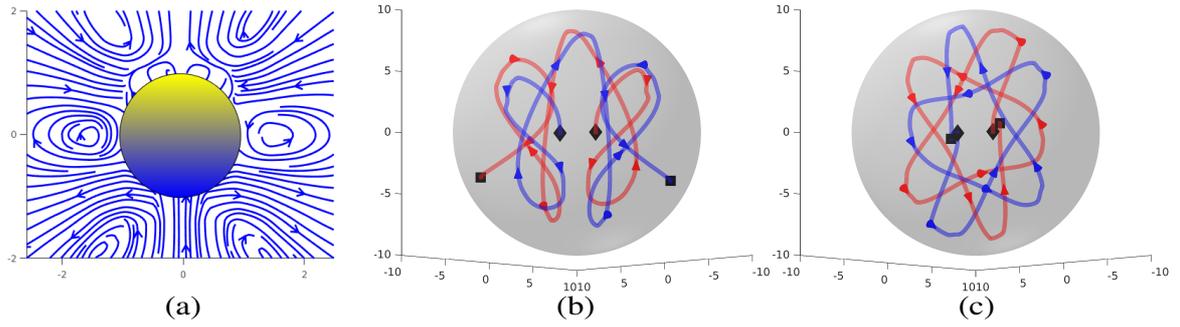}
    \caption{{\em We plot the fluid field of a single phoretic particle oriented in the positive y direction (a). In (b) and (c), two particles are placed in a shell of radius 10, with starting position denoted with a diamond. The particles are initially aligned in (b) and anti-aligned in (c). We plot the trajectories of the particle centers in the $xz$ plane.}}
    \label{fig:shell_trajectory}
\end{figure}

\noindent A single such particle in free space has speed $\frac{4}{45}a_k m_k$ in the direction of the particle head. Even for $p$ as low as four, we find that the velocity of a single particle in the simulations matches the theory with 15 digits of accuracy. We plot the flow generated from a single particle in Figure \ref{fig:shell_trajectory}, propelling the particle forward.

\paragraph{Pairwise interactions.}
We observe patterns of pairwise interactions when two particles are confined in a shell. In Figure \eqref{fig:shell_trajectory} we plot the trajectory of two particles in a symmetric orbit. We observe similar pairwise interaction behavior to that discussed in \cite{kanso2019phoretic}, which studied how interactions between particle pairs depended on their relative orientations, observing that anti-aligned particles tend to orbit each other. 

\paragraph{Many-body interactions.} Understanding the behavior of many-body phoretic suspensions is considerably more challenging. Previous studies of these systems tend to make a number of generalizations, such as assuming $A_k$ and $M_k $ to be constant on each hemisphere, or assuming the domain to be quasi-two dimensional and semi-infinite, as in \cite{kanso2019phoretic}.
In our case, we set the confining geometry $\Omega_\infty$ to be a sphere. We set the flux on the boundary to be such that the total flux on the system is 0:
\begin{equation}
\frac{dC}{d\nu} = \frac{1}{4\pi R_\infty^2}\sum_{k=1}^M\int_{\Gamma_k}A_k(\theta_k).
\end{equation}

\noindent With this configuration, we observe that the particles are attracted to the boundary of the shell, orienting themselves in the direction normal to the shell. This particle migration occurs rapidly.

\begin{figure}[H]
	\centering
\includegraphics[width = \linewidth]{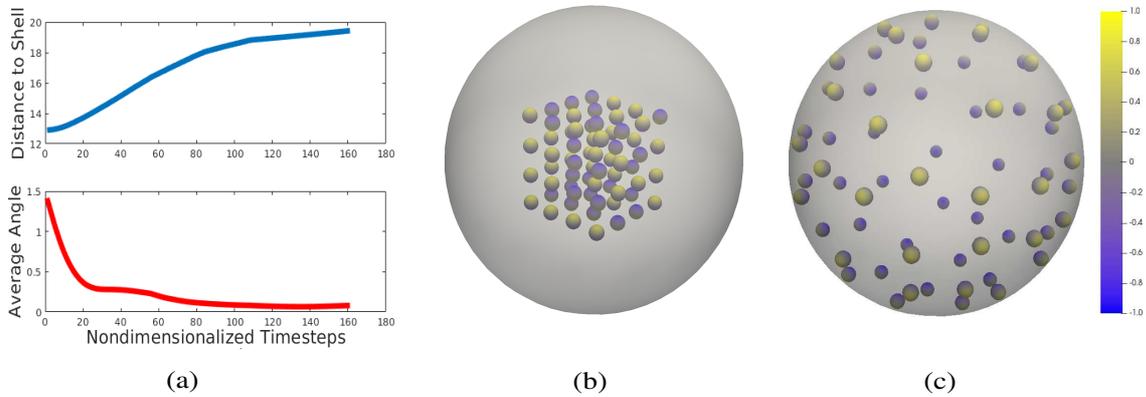}
	\caption{{\em We place a $4 \times 4 \times 4$ grid of phoretic particles in a confining shell of radius $25$. (b) The particles spread out and migrate to the shell, orienting themselves in the direction of the shell normal (c). The plots in (a) show the average distance between the particles and the shell and the average alignment between the particles and the shell normal at the point of contact.}
	\label{fig:diagonal}}
\end{figure}

%% file: conclusions.tex
\label{sec:conclusions}

We presented a general computational framework for the simulation of dense Janus particle suspensions in Stokes flow. Our approach features integral representations of long-range Janus particle interactions; for this purpose, we have contributed efficient and spectrally-accurate scalar potential evaluation methods for screened Laplace potentials. To resolve resulting fluid flow and particle collisions, we leverage recent developments in fast algorithms for high-fidelity Stokes rigid body problems.

All numerical solvers proposed for this framework are spectrally-accurate, efficient and scalable with problem size. Due to the favorable conditioning of the BIEs involved and our use of efficient evaluation schemes for both near-field and far-field spherical particle interactions, computational cost scales linearly with the number of particles. We note that, provided efficient singular and near-singular integral evaluation schemes are developed, it can be readily applied to wide classes of particle shapes and confining geometries. We are currently investigating the extent to which the spectral analysis techniques we have described can be extended to spheroidal, ellipsoidal and axisymmetric shapes.

Due to the nature of the physical fields involved in most relevant Janus particle types, the approach presented in this work has wide applicability. We demonstrate this through three distinct case studies of Janus particles of great relevance to applications in biomedicine and materials science: amphiphilic, bipolar electric, and phoretic particles. We note these examples do not constitute an exhaustive list; a number of additional Janus particle systems can be modeled by following the process outlined in this work. For instance, the bipolar formulation presented in this work may be readily adapted to problems in magnetic Janus particle suspension simulation \cite{seong2016magnetic,sobecki2021dynamics}. Moreover, the techniques presented here may be relevant to a larger class of active matter systems, for example, in the simulation of chemotactic bacterial suspensions. 

In each of these studies, we show how to design integral representations for the Janus interaction potential, leading to well-conditioned second kind boundary integral equations; depending on the coupling between Janus and hydrodynamic interactions, the corresponding integral-equation based Stokes rigid body solver is deployed. The ability to accurately simulate these suspensions allows us to recreate spontaneous self-assembly for moderately large systems of particles in a single processor; our experience with hybrid HPC implementations such as \cite{yan2019scalable} suggests the methods proposed here could be readily scaled to enable large-scale simulations on distributed-memory machines.

%% file: Appendix.tex
\section{Derivation of the Layer Spectra}

\label{appendix:Spectra}

In Theorem \ref{thm:LayerSpec}, we presented formulas for the spectra of $\SL$ and $\DL$ were presented. We outline the derivation of these values below, following the presentation in \cite{vico2014boundary}, where the spectra of the single and double layer potentials for the Laplace operator were derived. Here we derive the layer operator spectra for a single particle of radius $1.$

Let potential $\phi$ be a solution to the screened Laplace equation with parameter $\lambda.$ On the surface of a sphere, $\phi$ may be written as a superposition of spherical harmonics, so we make the ansatz that 
\[ \phi = a_n^m f_n(r)Y_n^m(\theta,\phi)\] 

\noindent for all $r.$
Plugging $\phi$ into the screened Laplace operator and employing orthogonality of the $Y_n^m,$ we obtain an ODE for $f_n(r):$ 
\begin{equation}
r^2 f^{''}_n + 2r f^{'}_n - \left[ \lambda^2 r^2 + n(n+1) \right] f_n = 0.
\end{equation}
This equation is known as the \textit{spherical Bessel differential equation} and has two sets of admissible solutions, called modified spherical Bessel and Hankel functions and denoted respectively by $i_n$ and $k_n$. These functions can be expressed in terms of modified Bessel functions as:

\begin{equation}
 i_n (\lambda r) = \sqrt{\frac{\pi}{2 \lambda r}} I_{n+\frac{1}{2}} (\lambda r),\quad k_n(\lambda r) =  \sqrt{\frac{\pi}{2 \lambda r}} K_{n+\frac{1}{2}} (\lambda r),     
\end{equation}
\noindent where $I_n(r)$ is the modified Bessel function of the first kind and $K_n(r)$ is the modified Bessel function of the second kind.  

\paragraph{Layer Potential.}
We use our representation of solutions to the screened Laplace equation in conjunction with properties of layer operators to solve for the spectral values of the single and double layer operators.

Let $\varphi = \SL_\lambda [Y_n^m].$ We have that  

\[ \varphi (r, \theta, \phi) =  \begin{cases} 
 \sum_{n=0}^\infty \sum_{m=-n}^n a_{nm}^o k_n(\lambda r) Y_n ^m (\theta, \phi) & r > 1 \\
 \sum_{n=0}^\infty \sum_{m=-n}^n a_{nm}^i i_n(\lambda r) Y_n ^m (\theta, \phi) & r < 1,
\end{cases}
\]

\noindent with $\{a_{nm}^o\}, \{a_{nm}^i\}$ unknown coefficients for the exterior and interior respectively. By using the continuity of $\SL$ at the boundary and orthogonality of $Y_n^m,$ we obtain an equation relating the coefficients:
\begin{subequations}

\begin{equation}
a_{nm}^o k_n (\lambda) - a_{nm}^i i_n (\lambda) = 0. 
\end{equation}

\noindent the jump condition $[[S'\phi]] = -\phi$ yields a second equation for the coefficients:
\begin{equation}
a_{nm}^o \lambda k'_n  (\lambda ) - a_{nm}^i \lambda i'_n (\lambda)  =  -1.
\end{equation}
\end{subequations}
Solving the $2 \times 2$ linear system for each pair of $(n,m)$ gives us that:
\begin{equation}
\centering
\begin{aligned}
   a_{nm}^i = \frac{- k_n(\lambda)}{\lambda \; W\left(i_n(\lambda), k_n(\lambda)\right)}, \quad & a_{nm}^o = \frac{- i_n(\lambda)}{\lambda \; W\left(i_n(\lambda), k_n(\lambda)\right)}, \\
\end{aligned}
\end{equation}
where $W$ is the Wronskian 

\[ W\left(i_n(\lambda), k_n(\lambda) \right) =  i_n(\lambda)k'_n(\lambda) - k_n(\lambda)i'_n(\lambda). \] 

\noindent It can be shown that $W\left(i_n(\lambda), k_n(\lambda)\right) = - \dfrac{\pi}{2 \lambda^2}$. From this, the values in \ref{thm:LayerSpec} follow. A nearly identical analysis yields the double layer coefficients, with only the jump conditions on the layer operator changing.

\section{Derivatives of Operators}
\label{appendix:Derivatives}

The formulas for the kernels of the normal derivatives of the layer operators $\SL$ and $\DL$ are given below. Here $\boldsymbol{\nu_x}$ and $\boldsymbol{\nu_y}$ are the normal derivatives with respect to $\boldsymbol{x}$ and $\boldsymbol{y}$ respectively. Also, let ${\zeta_x}$ denote the dot product ${\zeta_x} = \boldsymbol{\nu_x}^T \boldsymbol{(x-y)}$, and ${\zeta_y} = \boldsymbol{\nu_y}^T \boldsymbol{(x-y)}$.

\begin{subequations} 
\begin{equation}
    \SL'_\lambda=\frac{{\zeta_x}}{4\pi} \frac{e^{-\lambda \|\boldsymbol{x-y}\|}}{\|\boldsymbol{x-y}\|^2} \left( \frac{1}{\|\boldsymbol{x-y}\|}+\lambda \right),
\end{equation} 

\begin{equation}
\begin{aligned}  
     \DL'_\lambda = \frac{e^{-\lambda\|\boldsymbol{x-y}\|}}{4\pi} \bigg[ \frac{1}{\|\boldsymbol{x-y}\|^3} \left( \lambda^2 + \frac{2\lambda}{\|\boldsymbol{x-y}\|}+\frac{2}{\|\boldsymbol{x-y}\|^2} \right) {\zeta_x}{\zeta_y} \\
    - \left( \frac{\lambda}{\|\boldsymbol{x-y}\|} + \frac{1}{\|\boldsymbol{x-y}|^2} \right) \left( \frac{1}{\|\boldsymbol{x-y}\|}(\boldsymbol{v_x}^T\boldsymbol{v_y}) -  \frac{1}{\|\boldsymbol{x-y}\|^3}{\zeta_x}{\zeta_y} \right) \bigg].
\end{aligned}
\end{equation}
\end{subequations}
The spectra of the derivative operators are as follows: 

\begin{center}
$\begin{array}{c|c|c}
&\SL'_\lambda [Y_n^m] (r,\theta,\phi) & \DL'_\lambda [Y_n^m] (r,\theta,\phi) \\[2mm]
\hline 
r > 1 & \dfrac{2\lambda^2 i_n(\lambda)}{\pi} k_n'(\lambda r) Y_n ^m (\theta,\phi) & \dfrac{2\lambda^3 i'_n(\lambda)}{\pi} k_n'(\lambda r) Y_n ^m (\theta,\phi)\\[2mm]
\hline 
r < 1 & \dfrac{2\lambda^2 k_n(\lambda)}{\pi} i_n'(\lambda r) Y_n ^m (\theta,\phi) & \dfrac{2\lambda^3 k'_n(\lambda)}{\pi} i_n'(\lambda r) Y_n ^m (\theta,\phi) \\[2mm]
\end{array}.$ 
\end{center}
\section{Scaling analysis}
\label{appendix:scaling}

Throughout our analysis of the screened Laplace BIOs, we have defined them on the unit sphere. This is sufficient for calculations involving spheres of any size, as the following result holds:

\begin{lemma}
\label{lemma:scaling}
Let $\SL^r_\lambda$ and $\DL^r_\lambda$ be the single and double layer operators for the screened Laplace equation on the surface of a sphere of radius $r$ centered at the origin. Then,

\begin{enumerate}
\begin{minipage}{0.4\linewidth}    

\item $\SL^r_\lambda [\mu] (\boldsymbol{x}) = r \SL_{\lambda r} [\mu] (\boldsymbol{x} / r),$ 
    \item $\DL^r_\lambda [\mu](\boldsymbol{x}) = \DL_{\lambda r} [\mu ](\boldsymbol{x} / r) ,$ 
\end{minipage}
\begin{minipage}{0.4\linewidth}   

\item $\SL^{r \prime}_\lambda [\mu](\boldsymbol{x}) = \SL'_{\lambda r} [\mu ](\boldsymbol{x} / r), $
    \item $\DL^{r \prime}_\lambda [\mu](\boldsymbol{x}) = \frac{1}{r} \DL'_{\lambda r} [\mu](\boldsymbol{x} / r).$ 
\end{minipage}

\end{enumerate}
\end{lemma}

These properties can easily be verified by a simple change of variables procedure (e.g., $y' = r y$) to the integrals \eqref{eqn:SreenedLap}. Given a routine that evaluates the quantities in \ref{thm:LayerSpec}, the above lemma allows for the same quantities to be evaluated on spheres of any size by scaling the input and output and changing parameter $\lambda$ to $\lambda r$.

\section{Nondimensionalization}
\label{app:Nondim}
We discuss the nondimensionalization of units in physical applications. In sections \ref{sec:amphiphillic} and \ref{sec:Bipolar}, a natural choice of unit length, $L$ is the Debye length, $\lambda^{-1}.$

\paragraph{Amphiphilic Janus particles.} 
In addition to the characteristic length $L=\lambda^{-1},$ we define a characteristic value of the potential, $\phi_c.$ Prior to nondimensionalization, the hydrophobic stress tensor is given by: 

\begin{equation}
T_H =  \left(\lambda^2 \phi^2 I + 2(\|\nabla\phi\|^2 - \nabla\phi\otimes \nabla\phi)\right). 
\end{equation}
We nondimensionalize this expression by substituting $\frac{\phi}{\phi_c}$ into the tensor, which yields 

\begin{equation}
    T_H = \frac{\phi_c^2}{L^2} T_H'.
\end{equation}
$\frac{\phi_c^2}{L^2}$ has units of pressure. We refer to it as the \textit{amphiphilic pressure}, $\pi_a$. We follow the standard nondimensionalization of the Stokes equation:
\begin{equation}
T_S = \frac{\mu u_c}{L_c} ( -p_c I + \left(\nabla u' + \nabla u'^T) \right) = \pi_v T_S',
\end{equation}

\noindent where $u_c$ is a characteristic fluid speed and $\pi_v = \frac{\mu u_c}{L}$ is referred to as the \textit{viscous pressure}.
Equating the two tensors and dividing through by the viscous pressure, we obtain 
\[ T_S = \eta T_H, \]
where $\eta = \pi_a / \pi_v$ is the ratio of amphiphilic pressure to viscous pressure. 

\paragraph{Bipolar particles.} 
We again let $L$ equal the Debye length. Since we have normalized the exterior screened Laplace equation, $L=1$. In this case, we use the potential from the electric field to define $\phi$ in terms of $\|\boldsymbol{E_0}\|L.$. Using these values in the Maxwell Stress tensor, we obtain 
 
\begin{equation}
 T_E =  (\epsilon_0) \|\boldsymbol{E_0}\|^2 \left(\nabla\phi' \times \nabla\phi' - \frac{\|\nabla\phi'\|^2}{2}\right).
\end{equation}
$\epsilon_0 \|\boldsymbol{E_0}\|^2$ is two times the \textit{electrostatic pressure} and may be denoted as $\pi_e$. 
Just like in the amphiphilic case, we can equate this tensor with the Stokes tensor and define $\eta$ as the ratio between electrostatic and viscous pressures yielding 
\begin{equation}
T_S = \eta T_E,
\end{equation}
where $\eta = \pi_e / \pi_v$.

\paragraph{Phoretic particles.}
We model phoretic particles with the Laplace equation, so we cannot use a Debye length as the unit length. Rather, we let $L$ be the particle radius.

The concentration is modelled by a diffusion equation:
\begin{equation}
    D\nabla^2 C = 0.
\end{equation}
\noindent We set the unit time to be $T=L^{-1/2}$ so that $D$ is dimensionless.

Unlike the amphiphilic and bipolar cases, the key coupling between Janus and hydrodynamic interactions for phoretic particles occurs through the tangential slip velocity induced at particle boundaries. After nondimensionalization, we obtain

\begin{equation}
    \boldsymbol{u}_{slip} = \frac{D}{u_c L} M_{k}(\theta_k)(I-\boldsymbol{n}\boldsymbol{n}^T)\nabla C,
\end{equation}

\noindent where $C_c$ is the unit concentration. The quantity $(D/L)/u_c$ is a ratio between the speed of diffusion and the fluid speed. For all experiments presented in this work, we choose $u_c$ to be the speed of a single phoretic particle in unbounded flow. 